\documentstyle[aps,preprint,prl]{revtex}

\begin{document}

\bibliographystyle{prsty}
\draft

\title{Coulomb interaction and ferroelectric instability of BaTiO$_{3}$}
\author{Ph. Ghosez, X. Gonze and J.-P. Michenaud}
\address{Unit\'e de Physico-Chimie et de Physique des Mat\'eriaux,
Universit\'e Catholique de Louvain,\\
1 Place Croix du Sud, B-1348 Louvain-la-Neuve, Belgium}
\date{\today}
\maketitle
\begin{abstract}
Using first-principles calculations, the  phonon frequencies at the
$\Gamma$ point and the
dielectric tensor are determined and analysed for the cubic and
rhombohedral phases of  BaTiO$_{3}$.
The dipole-dipole interaction is then separated \`a la Cochran from the
remaining short-range forces,
in order to investigate their respective influence on lattice dynamics.
This analysis highlights the
delicate balance of  forces leading to an unstable phonon in the cubic
phase and demonstrates its
extreme sensitivity to effective charge changes. Within our decomposition,
the stabilization of the
unstable mode in the rhombohedral phase or under isotropic pressure has a
different origin.

\bigbreak

dynamics: general theory).
\end{abstract}
\setcounter{page}{1}

\newpage


Barium Titanate (BaTiO$_{3}$) is a typical ferroelectric material that
undergoes three
temperature phase transitions, from a paraelectric cubic phase, stable at high
temperature, to ferroelectric phases of tetragonal, orthorhombic and
rhombohedral symmetry.
There have been considerable efforts to identify the
origin of the transitions in this particular ABO$_{3}$ compound
\cite{Lines}. Among them, let us point out the seminal theory of Cochran
\cite{Cochran}.
In the framework of a shell model, he relates the ferroelectric transition
to the softening of a
transverse optic phonon at $\Gamma$. Within his model, the interatomic
forces are decomposed into
short-range forces  and long-range Coulomb (dipole-dipole)  interaction.
The latter is evaluated  by
means of a Lorentz effective electric field, assuming a local spherical
symmetry at each atomic site.
Interestingly, the decomposition isolates the contribution of each kind of
forces on the
frequency of the transverse modes and identifies the structural instability
with the
cancellation of the two contributions. Although meaningful,  Cochran's
model is only  qualitative. The
numerical investigation is subject to many approximations. Moreover, it was
shown by
Slater\cite{Slater} that the Lorentz field is far from spherical.

In subsequent studies, it has been usually accepted that ferroelectricity
in perovskites results
from a delicate balance between short-range repulsions which favor the
cubic phase and
long-range electrostatic forces which favor the ferroelectric state.
Although some calculations
\cite{Cohen-KSV} illustrate this picture, they do not  rely on a well
defined separation
of the interactions.

In contrast, a separation of the interatomic forces was proposed recently
by Gonze  {\it et
al.}~\cite{Gonze-IFC}. Without postulating any atomic site symmetry, they
introduce an analytic form
for the dipole-dipole interaction at the microscopic level from Born
effective charges and dielectric
tensor.  This formulation, evaluated thanks to first-principles data,
generalizes Slater's calculation of
the Lorentz field  \cite{Slater} and can be used to refine Cochran's
results \cite{Cochran}.

In our letter, we consider the cubic and
rhombohedral phases as well as a  compressed cubic structure.  We first
report dielectric tensor
values,  then compute the  dynamical matrices and phonon frequencies at the
$\Gamma$ point and
compare them with experiment. Separating a dipole-dipole interaction from
the short-range
remaining part of the dynamical matrix following Gonze {\it et
al.}~\cite{Gonze-IFC},  we
quantify the balance of forces generating the unstable cubic phonon mode and
investigate its sensitivity to effective charge changes. Within this
decomposition,  the hardening of
the unstable phonon in  the rhombohedral  phase or under isotropic pressure
has a different origin.


Computational details are the same as in Ref.~\cite{Lee,GGM2,RC}:
Calculations are performed within
the Density Functional Theory (DFT) and the Local Density Approximation
(LDA)  using a
conjugate-gradient plane-wave pseudopotential method.  Responses to
electric field and
phonon-type perturbations are obtained within a variational approach to
Density Functional
Perturbation Theory.


We consider cubic structures at the experimental and theoretically
optimized volumes
with lattice parameters $a_{o}$ = 4.00 and 3.94~\AA,
as well as a compressed cubic cell with $a_{o}$=3.67~\AA. For the
rhombohedral phase,
we worked at the experimental unit cell parameters, with relaxed
atomic positions. Accurate values for the Born effective charges
$Z^{*}_{\kappa,\alpha\beta}$
were already reported elsewhere \cite{RC}.

The computed dielectric tensor $\epsilon_{\infty}$ is presented in Table
\ref{diel} in comparison to
its experimental estimate in the cubic phase \cite{Burns}.  No previous values
were reported for the rhombohedral phase.  It is well known that the
DFT-LDA usually overestimates
the experimental $\epsilon_{\infty}$. For the cubic geometry, the
discrepancy  is of the order of 25\%.
This error can be corrected in first approximation  by the {\it scissor
operator} technique \cite{SCI}.
For all cases, we have used a scissor shift of $1.36\,eV$ that adjusts the
bandgap at the $\Gamma$
point in the experimental cubic structure to the value of $3.2\,eV$
\cite{Lines}.
For the cubic phase, the scissor corrected $\epsilon_{\infty}$ (5.61)
overestimates  the
experimental value (5.40) by less than 5\%. For the rhombohedral structure,
the values are
globally smaller, especially along  the ferroelectric axis. This goes
hand-in-hand with the reduction
also observed in this direction for $Z^*_{\kappa}$ \cite{RC}.

There are 12 optical phonons in BaTiO$_{3}$.
In the cubic phase, we have three triply degenerate modes of $F_{1u}$
symmetry and a triply degenerate mode of $F_{2u}$ symmetry.
Only the $F_{1u}$ modes are infrared active, with  LO-TO splitting,
while the $F_{2u}$ modes are {\it silent} modes that cannot be identified
experimentally. Our values (Table \ref{phonons}) at the
optimized volume are in close agreement with the experiment \cite{Luspin}
as the theoretical
results of Zhong {\it et al.}\cite{Zhong-Z}. In particular,  we reproduce
the instability
of the TO1 mode corresponding to the  vibration of Ti and Ba against the
O atoms. The phonon frequencies appear very sensitive to the small  volume
change
from the experimental to the theoretical cubic phase,  contrary to
Z$^*_{\kappa}$ \cite{RC} or $\epsilon_{\infty}$.  This is particularly true
for the soft TO1
mode, whose instability even disappears in our compressed cubic phase.

Due to the long-range Coulomb interaction, the eigendisplacements
of the TO modes ($\eta^{TO}$) do not necessarily correspond to those of the
LO modes ($\eta^{LO}$).
The overlap matrix $\langle \eta^{TO}|M|\eta^{LO} \rangle$  reported in
Table \ref{overlap}
establishes however that the mixing of modes is very weak in the cubic
phase ( $M$ is such that
$M=M_{\kappa} \delta_{\kappa \kappa'}$ with $M_{\kappa}$ is the mass of
atom $\kappa$).
In agreement with this observation, assuming that $\eta^{LO}$ and
$\eta^{TO}$ are identical, the fictitious LO frequencies predicted  on the
basis  of the oscillator
strengths (Eq. 10 of Ref. \cite{Lee}) are respectively of 701, 214 and 508
cm$^{-1}$, in close
agreement with the theoretical LO frequencies. Note the giant splitting of
the TO1 mode
already mentioned by  Zhong {\it et al.}\cite{Zhong-Z}.
It arises from the large effective charges on Ti ($Z^*_{Ti}=+7.28$) and O
($Z^{*}_{O_{\parallel}}=-5.73$, for a displacement along the Ti-O bond)
generating a mode effective
charge
$Z^{*}_{TO1}= \left\|  \frac{\sum_{\kappa,\beta} Z^{*}_{\kappa,\alpha \beta} \,
\eta^{TO1}_{\kappa,\beta}} {\langle \eta^{TO1} | \eta^{TO1} \rangle}
\right\| =9.02$.

In the rhombohedral phase (Table \ref{phonons}), each triply
degenerate $F_{1u}$ (resp. $F_{2u}$) mode from the cubic phase gives
rise to a mode of $A_{1}$ (resp. $A_{2}$) symmetry with
eigendisplacements along the ferroelectric direction, and a doubly degenerate
mode of $E$ symmetry. $E$ and $A_{1}$ modes are infrared and Raman active.
The only relevant comparative result we found is experimental
\cite{Laabidi} and  localizes
the phonon frequencies in three regions (100-300 cm$^{-1}$,
480-580 cm$^{-1}$, and 680-750 cm$^{-1}$)  in qualitative
agreement with our values.

All the modes are stable in the rhombohedral structure.
Due to the small distortions, the eigenvectors remain very similar to those
of the cubic phase.  Table
\ref{overlap} compares $A_{1}$ to corresponding $F_{1u}$ eigenvectors.
Similar  values are obtained
for the $E$ modes. They point out that  $A_{1}(TO2)$ and $E(TO2)$ originate
from the hardening of
the soft mode. Even if both of these modes continue to couple strongly with
the electric field, the
smaller $Z^*_{\kappa}$ make their mode effective charge smaller: 7.00 and
8.41 respectively.
We predict a static dielectric constant equal to 33.09  along the
ferroelectric direction and  to 68.89
perpendicularly to it.

The phonon frequencies $\omega$  and the
associated eigendisplacements $\eta$ are deduced from the dynamical matrix $A$
through the following equation:
$
\sum_{\kappa' \beta} A_{\alpha \beta}(\kappa \, \kappa')  \,
\eta_{\kappa',\beta} = M_{\kappa} \,
\omega^2 \, \eta_{\kappa,\alpha}
.$
The $\alpha$ and $\beta$ indices denote the space direction while $\kappa$
and $\kappa'$ label the atom within the unit cell. The ansatz proposed  by
Gonze  {\it et
al.}~\cite{Gonze-IFC} can now be  used to parametrize the  dipole-dipole
contribution to the
interatomic force constant from the knowledge of $Z^{*}_{\kappa}$ and
$\epsilon_{\infty}$,  in the general case where these tensors are
anisotropic \cite{Rem1}:
\begin{eqnarray*}
\Phi_{\alpha \beta}^{DD}(0\kappa,j\kappa')=
\sum_{\alpha' \beta'} Z^{*}_{\kappa,\alpha\alpha'} \,
Z^{*}_{\kappa',\beta\beta'} \,
(det \, \epsilon_{\infty})^{-\frac{1}{2}}
\left( \frac{(\epsilon^{-1}_{\infty})_{\alpha' \beta'}}{D^3}
-3 \frac{\Delta_{\alpha'}\Delta_{\beta'}}{D^5} \right)
\end{eqnarray*}
where $\Delta_{\alpha}=\sum_{\beta}(\epsilon^{-1}_{\infty})_{\alpha \beta}
\, d_{\beta}$,
$\vec{d}=\vec{R}_{j} + \vec{\tau}_{\kappa'} - \vec{\tau}_{\kappa}$ is the
vector relating nuclei , and
$D=\sqrt{\vec{\Delta}.\vec{d}}$. The contribution of this dipole-dipole
term to the
dynamical matrix is evaluated using  Ewald summation
technique~\cite{Gonze-IFC}.  By this way,
dipole-dipole ($DD$) and remaining short-range ($SR$) \cite{Rem2} parts of
the dynamical matrix
$A$ can be isolated from each other {\it \`{a} la Cochran}
($A=A_{DD}+A_{SR}$) and their partial
contribution to $\omega^2$ can be evaluated as follows:
$$
\underbrace{\langle \eta | A | \eta \rangle}_{\omega^2}
= \underbrace{\langle \eta | A_{DD} | \eta \rangle}_{\omega^2_{DD}}
+ \underbrace{\langle \eta | A_{SR} | \eta \rangle}_{\omega^2_{SR}}.
$$
$A_{DD}$ and $A_{SR}$ can then
be modified independently in order to investigate their own influence on
the instable mode.

In Table \ref{ddsr-cubic} we report the values of $\omega^2_{DD}$ and
$\omega^2_{SR}$ for the TO
modes of the cubic phase at the optimized volume. We observe that the
instability of the
$F_{1u}(TO1)$ mode originates from the compensation of two very large
numbers, the $DD$
interaction greatly destabilizing the crystal. Interestingly, this close
compensation exists
for the unstable mode {\it only}.

In the cubic phase, the large values of  $Z^{*}_{Ti}$ and
$Z^{*}_{O_{\parallel}}$ (responsible of the
strong Coulomb interaction) are  mainly produced by a dynamic transfer of
charge along the
Ti-O bond \cite{RC}.  Postulating $A_{SR}$ to be fixed, we can fictitiously
reduce this transfer  of
charge by decreasing simultaneously $Z^{*}_{Ti}$ and
$Z^{*}_{O_{\parallel}}$,  and monitor the $F_{1u}(TO1)$ mode frequency
changes. Figure 1 shows
that $\omega^{2}(TO1)$ evolves quasi linearly with the transfer of charge.
A change
corresponding to a reduction of the order of 1\% of $Z^*_{Ti}$ is enough
to {\it suppress} the
instability.  Of course  this situation is artificial: In a real material
any modification of
$Z^{*}_{\kappa}$ would go  hand-in-hand with a change of the SR forces.
This result
however highlights the very delicate compensation existing between $DD$ and
$SR$  interactions.
Interestingly, $\omega^2_{SR}$ is also modified, due to the change of the
eigenvector $\eta$ induced
by the modification of $A_{DD}$. This change is not crucial and a similar
evolution of
$\omega^2$ is observed if we keep the eigenvector of the original optimized
structure.
Note that all these conclusions are independent of the use of the scissor
correction for
$\epsilon_{\infty}$. From now, we report results without scissor correction.

In the rhombohedral structure,  there is no unstable mode although
the eigenvectors remain close to  those of the cubic phase (see Table
\ref{overlap}).
It was found \cite{RC} that $Z^*_{\kappa}$  are smaller in this
ferroelectric phase, suggesting a
smaller DD interaction, but this could be partly compensated by a reduction
of $\epsilon_{\infty}$.
For the $A_1(TO2)$ mode coming from the soft mode, $\omega^2_{DD}$ (-286267
cm$^{-2}$) add to a slightly larger $SR$ counterpart (356373 cm$^{-2}$).
The values
differ widely from those of the cubic phase: The
$SR$ forces give less stabilization but this is compensated by a larger
reduction of the $DD$ contribution.

If we now fictively modify $A_{DD}$ and replace $Z^*_{\kappa}$ and
$\epsilon_{\infty}$  of the
ferroelectric structure by their value in the cubic phase, we modify the
frequency of the
$A_1(TO2)$ mode from 265 to 266$i$ cm$^{-1}$: We obtain an instability even
{\it larger} than in the
cubic phase. From this point of view,  the reduction of $Z^*_{\kappa}$ in
the rhombohedral
phase appears as a crucial element to the  stabilization of the
$A_1(TO2)$ mode. Introducing $Z^*_{\kappa}$ and $\epsilon_{\infty}$ of the
cubic phase,
$\omega^2_{DD}$ and $\omega^2_{SR}$  are also strongly  modified  and
becomes respectively
-871017 and 800371 cm$^{-2}$.  The dramatic change of
$\omega^2_{SR}$ results from a change of eigenvector pointing out the
anisotropy of the $SR$
forces (the overlap between the new and original eigenvector is equal to
0.86). If we had kept the
eigenvector unchanged,  we would still have observed an instability (74$i$
cm$^{-1}$) for the
$A_1(TO2)$ mode although much smaller. This means that the inclusion of the
effective charges of the
cubic phase is already sufficient to destabilize the crystal but at the
same time produces a change of
eigenvector enlarging the instability.

No more instability is present in the compressed cubic phase, although the
global values of
$Z^*_{\kappa}$ do not differ significantly from those at the optimized
volume\cite{RC}.
Moreover, the reduction of volume even increases the destabilizing effect
of the $DD$ interaction by
20\%. However,  strong modifications of the $SR$ forces produce a mixing of
modes so that no
one can be identified with the unstable mode observed at the optimized
volume.  If we replace
$A_{SR}$ by its value at the optimized volume we recover a very large
instability (437$i$
cm$^{-1}$). The disappearance of the  unstable mode under pressure seems
therefore essentially
connected to a modification of the $SR$ forces in contrast to its
stabilization in the rhombohedral
phase related to the reduction of $Z^*_{\kappa}$.

We thank
Corning Inc. for the availability of the {\sc planewave} code,
and
J.-M. Beuken for computer assistance.
X.G. is grateful to FNRS-Belgium for financial support.
We used  IBM-RS6000  from common projects between IBM Belgium,
UCL-PCPM  and FUNDP.


\begin{figure}
\caption{Evolution of the $F_{1u}(TO1)$ mode frequency squared with respect
to the dynamic transfer
of charge along  the Ti-O bond (quantified here by the evolution of
$Z^*_{Ti}$, see text). Results are
obtained with ($\circ$)  or without (   $\bullet$ \hspace{-.18in}---)
scissor shift for $\epsilon_{\infty}$.
SR and DD contributions to $\omega^2$ (see text) are shown in the inset. }
\end{figure}

\begin{table}
\caption {Dielectric tensor of BaTiO$_3$
obtained within the local density approximation (LDA) or
with an additional scissor correction (SCI).
For the rhombohedral phase, the $z$ axis points
in the ferroelectric direction.}
\label{diel}
\begin {tabular}{lccccccc}
& &\multicolumn{3}{c}{Cubic phase
($\epsilon_{\infty}^{xx}=\epsilon_{\infty}^{yy}=\epsilon_{\infty}^{zz}$)}
& &\multicolumn{2}{c}{Rhombohedral phase}\\
& &$a_{o}=3.67$ \AA &$a_{o}=3.94$ \AA &$a_{o}=4.00$ \AA
& &$\epsilon_{\infty}^{xx}=\epsilon_{\infty}^{yy}$ &$\epsilon_{\infty}^{zz}$ \\
\hline
LDA & &6.60  &6.66  &6.73 & &6.16  &5.69  \\
SCI  & &5.71  &5.60  &5.61 & &5.26  &4.91  \\
\end{tabular}
\end{table}

\begin{table}
\caption {Phonon frequencies (cm$^{-1}$)  at the $\Gamma$ point
for cubic and rhombohedral BaTiO$_{3}$.}
\label{phonons}
\begin {tabular}{lccccclclc}
\multicolumn{5}{c}{Cubic phase} & & \multicolumn{4}{c}{Rhombohedral phase}\\
Mode &$a_{o}$=3.67\AA &Exp.\cite{Luspin} &$a_{o}$=3.94\AA &$a_{o}$=4.00\AA  &
&Mode   &   &Mode   & \\
\hline
$F_{1u}(TO1)$   &214    &soft  &113$i$  &219$i$    & &$A_{1}(TO1)$ &168
&$E(TO1)$ &161\\
$F_{1u}(LO1)$   &250    &180   &180       &159     &     &$A_{1}(LO1)$ &180
&$E(LO1)$  &173\\
$F_{1u}(TO2)$   &296    &182  &184       &166      &    &$A_{1}(TO2)$  &265
&$E(TO2)$ &205\\
$F_{1u}(LO2)$   &513    &465  &460        &447     &    &$A_{1}(LO2)$  &462
&$E(LO2)$  &438\\
$F_{1u}(TO3)$   &737    &482  &481        &453     &   &$A_{1}(TO3)$  &505
&$E(TO3)$  &461\\
$F_{1u}(LO3)$    &1004  &710  &744        &696     &   &$A_{1}(LO3)$  &702
&$E(LO3)$   &725 \\
$F_{2u}$             &308    &306  &288        &281    &     &$A_{2} $
&274    &$E $ &293\\
\end{tabular}
\end{table}

\begin{table}
\caption {Overlap matrix elements between the eigenvectors of the
$F_{1u}(TO)$ modes
of the optimized cubic phase and those respectively of the associated
$F_{1u}(LO)$ mode
and  of the $A_{1}(TO)$ mode of the rhombohedral phase.}
\label{overlap}
\begin {tabular}{lcccccccccc}
& &
&$F_{1u}(LO1)$ &$F_{1u}(LO2)$ &$F_{1u}(LO3)$
& &
&$A_{1}(TO1)$ &$A_{1}(TO2)$ &$A_{1}(TO3)$\\
\hline
$F_{1u}(TO1)$ & &  &0.17  &-0.36 &0.92      & & &0.13 &-0.97 &-0.19\\
$F_{1u}(TO2)$ & &  &-0.99  &-0.07  &-0.16 & & &-0.99  &-0.13  &-0.01\\
$F_{1u}(TO3)$ & &  &0.01  &-0.93 &0.37      & & &-0.02 &0.18  &-0.98\\
\end{tabular}
\end{table}

\begin{table}
\caption { $DD$ and $SR$ contributions (see text) to the TO mode frequency
squared (cm$^{-2}$)  for
the cubic phase at the optimized volume. Values in brackets where obtained
with the
scissor-corrected value of $\epsilon_{\infty}$.}
\label{ddsr-cubic}
\begin {tabular}{lccccc}
&   &$F_{1u}(TO1)$    &$F_{1u}(TO2)$   &$F_{1u}(TO3)$  &$F_{2u}$  \\
\hline
$\omega^{2}_{DD}$ &  &-625897 (-745610)  &7232 (8615)  &-130549 (-155518)
&109745 (130736)\\
$\omega^{2}_{SR}$ &  &613107 (732820)  &26538 (25155)  &361998 (386967)
&-26951 (-47942)\\
\hline
$\omega^{2}$            &                &-12790 \ \ \ \ \ \ \ \ \ \ \ \
&33770 \ \ \ \ \ \ \ \ \ \ \
&231449 \ \ \ \ \ \ \ \ \ \ \ \  &82794 \ \ \ \ \ \ \ \ \ \ \   \\
\end{tabular}
\end{table}


\vspace{2mm}


\begin{thebibliography}{10}
\bibitem{Lines}
M. E. Lines and A.M. Glass,
{\it Principles and Applications of Ferroelectrics and Related Materials}
(Clarendon Press, Oxford, 1977).
\bibitem{Cochran}
W. Cochran,
Adv. in Phys.  9, 387 (1960).
\bibitem{Slater}
J. C. Slater,
Phys. Rev. 78, 748 (1950).
\bibitem{Cohen-KSV}
R. E. Cohen,
Nature 358, 136 (1992);
R.D. King-Smith and D. Vanderbilt,
Phys. Rev. B, 49 , 5828 (1994).
\bibitem{Gonze-IFC}
X. Gonze, J.-C. Charlier, D.C. Allan and M.P. Teter,
Phys. Rev. B 50, 13035 (1994).
\bibitem{Lee}
C. Lee, Ph. Ghosez and X. Gonze,
Phys. Rev. B 50, 13379 (1994).
\bibitem{GGM2}
Ph. Ghosez, X. Gonze and J.-P. Michenaud,
Ferroelectrics 164, 113 (1995).
\bibitem{RC}
Ph. Ghosez, X. Gonze, Ph. Lambin and J.-P. Michenaud,
Phys. Rev. B 51, 6765 (1995).
\bibitem{Burns}
$\epsilon_{\infty}$ was obtained by  extrapolating to zero frequency index of
refraction measurements [G. Burns and F. H. Dacol, Solid State Comm., 42, 9
(1982)]
at different wavelengths.
\bibitem{SCI}
Z.H. Levine and D. C. Allan,
Phys. Rev. Lett., 63, 1719 (1989);
X. Gonze, Ph. Ghosez and R. W. Godby,
Phys. Rev. Lett. 74, 4035 (1995).
\bibitem{Luspin}
Y. Luspin, J.-L. Servoin and F. Gervais,
J. Phys. C 13, 3761 (1980).
\bibitem{Zhong-Z}
W. Zhong, R. D. King-Smith and D. Vanderbilt,
Phys. Rev. Lett. 72, 3618 (1994).
\bibitem{Laabidi}
K. Laabidi, M. Fontana and B. Jannot,
Sol. St. Comm. 76, 765 (1990).
\bibitem{Rem1}
In this formula, the macroscopic $\epsilon_{\infty}$ is used to parametrize
dipole-dipole
interactions down to nearest neighbors. No correction for the q-dependence
of $\epsilon_{\infty}$
and $Z^*_{\kappa}$ is included. This procedure is the natural
generalization of Luttinger and Tisza
calculations [Phys. Rev. 70, 954 (1946)], basis of Slater evaluation of the
Lorentz field \cite{Slater}.
\bibitem{Rem2}
The SR part also contains higher Coulomb terms like dipole-octupole and
octupole-octupole
interactions.
\end{thebibliography}
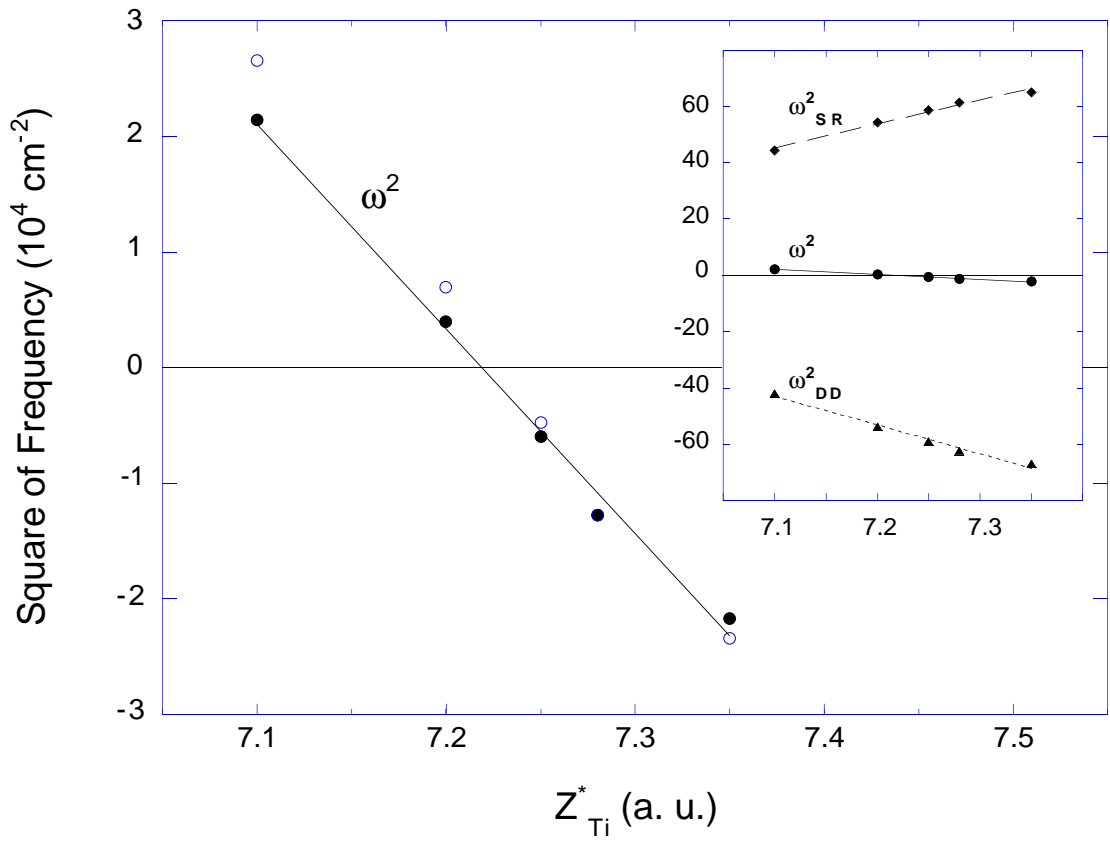
\end{document}